\documentstyle[preprint,aps]{revtex}
\newcommand{\beq}{\begin{equation}}
\newcommand{\eeq}{\end{equation}} 
\newcommand{\beqa}{\begin{eqnarray}}
\newcommand{\eeqa}{\end{eqnarray}} 
\newcommand{\ba}{\begin{array}} 
\newcommand{\ea}{\end{array}} 

\begin{document}
\draft


\widetext 
\title{Time-Dependent Variational Approach \\
to Bose-Einstein Condensation} 
\author{Luca Salasnich} 
\address{Istituto Nazionale per la Fisica della Materia, 
Unit\`a di Milano \\
Dipartimento di Fisica, Universit\`a di Milano \\
Via Celoria 16, 20133 Milano, Italy}
\maketitle
\begin{abstract}
We discuss the mean-field approximation for a trapped 
weakly-interacting Bose-Einstein 
condensate (BEC) and its connection with the exact many-body problem 
by deriving the Gross-Pitaevskii action of the condensate. 
The mechanics of the BEC in a harmonic potential 
is studied by using a variational 
approach with time-dependent Gaussian trial wave-functions. 
In particular, 
we analyze the static configurations, the stability 
and the collective oscillations for both ground-state and vortices.  
\end{abstract}

\vskip 1. truecm



\narrowtext

\newpage 

\section{Introduction}

Few years ago, the Bose-Einstein condensation 
has been observed with alkali-metal vapors 
confined in magnetic traps at very low temperatures.$^{1-3}$ 
By now, the total number of groups that have achived Bose-Einstein 
condensation is over a dozen. 
The confining potential is accurately described by a harmonic trap, 
and theoretical studies of the Bose-Einstein condensate (BEC) in 
harmonic traps have been performed for the ground state, 
elementary excitations and vortices (for a review see Ref. 4).  
\par 
In the first part of this paper, we discuss the time-dependent 
Gross-Pitaevskii (GP) equation,$^{5,6}$ which describes the macroscopic 
wave-function of a weakly-interacting Bose condensate at zero temperature. 
We deduce the GP action and the GP equation by imposing 
the least action principle to the quantum many-body action of the 
system within the mean-field approximation.$^{7}$ 
Such approach is mathematically equivalent to the field theory one, 
but it avoids the tricky issues related with spontaneous symmetry breaking. 
\par 
The rest of the paper is devoted to the study of BEC 
by using Gaussian trial wave-functions 
to minimize the GP action. 
Minimizing the GP action with a Gaussian trial function 
makes it possible to obtain approximate solutions for the condensate 
wave function, which can be treated, to a large extent, analytically. 
This method has been used by several authors 
to study the BEC static properties 
and the dynamics near the ground-state.$^{8-14}$ 
Here we extend its application also to vortex states. 
The presence of vortex states is a signature of the 
macroscopic phase coherence of the system. Moreover, 
vortices are important to characterize the superfluid properties 
of Bose systems.$^{7}$ 

\section{Variational Principle and Mean-Field approximation} 

Let us consider a $N$-body quantum system with 
Hamiltonian ${\hat H}$. The exact time-dependent Schr\"odinger equation 
can be obtained by imposing the quantum least action principle to the action 
\beq 
S= \int dt <\psi (t) | i\hbar 
{\partial \over \partial t} - {\hat H} |\psi (t) > \; ,
\eeq       
where $\psi$ is the many-body wave-function of the system and 
\beq
<\psi (t) |{\hat A} |\psi (t) > 
= \int \; d^3{\bf r}_1 ... d^3{\bf r}_N\;  
\psi^*({\bf r}_1,...,{\bf r}_N,t) {\hat A} 
\psi ({\bf r}_1,...,{\bf r}_N,t) \; , 
\eeq
for any quantum operator ${\hat A}$. 
Looking for stationary points of $S$ with respect 
to variation of the conjugate wave-function $\psi^*$ gives 
\beq
i\hbar {\partial \over \partial t}\psi = {\hat H}\psi \; , 
\eeq 
which is the many-body time-dependent Schr\"odinger equation. 
\par
As is well known, except for integrable systems, 
it is impossible to obtain the exact solution 
of the many-body Schr\"odinger equation and some approximation must be used. 
\par
In the mean-field approximation the total wave-function 
is assumed to be composed of independent particles, i.e. it can be 
written as a product of single-particle wave-functions $\phi_j$. 
In the case of identical fermions, $\psi$ must be antisymmetrized.$^{7}$ 
By looking for stationary action with respect to variation of a 
particular single-particle conjugate wave-function $\phi_j^*$ one finds 
a time-dependent Hartree-Fock equation for each $\phi_j$: 
\beq
i\hbar {\partial \over \partial t}\phi_j = {\delta \over \delta \phi_j^*} 
<\psi | {\hat H}| \psi > = {\hat h} \phi_j \; ,
\eeq
where ${\hat h}$ is a one-body operator. 
In general, the one-body operator ${\hat h}$ is nonlinear. Thus 
the Hartree-Fock equations are non-linear integro-differential 
equations. Note that for alkali-metal vapors, which are quite 
diluite, the independent-particle approximation gives 
reliable results.$^{4}$ 
Obviously, it is possible to go beyond the independent-particle 
approximation by including correlations in the many-body 
wave-function.$^{8}$ This is particularly useful to reproduce 
the properties of the superfluid $^4$He, that is a strongly 
interacting system.$^{9,10}$ 

\section{Gross-Pitaevskii Action}

In this section we discuss the zero-temperature mean-field approximation 
for a system of trapped weakly-interacting 
bosons in the same quantum state, i.e. a Bose-Einstein condensate.$^{7}$ 
In this case the Hartree-Fock equations reduce to only one equation, 
the Gross-Pitaevskii equation,$^{5,6}$ which describes the dynamics 
of the condensate. As previously discussed, 
this equation is intensively studied 
because of the recent experimental achievement of Bose-Einstein 
condensation for atomic gasses 
in magnetic traps at very low temperatures.$^{1-3}$ 
\par
The Hamiltonian operator of a system of $N$ identical 
bosons of mass $m$ is given by 
\beq
{\hat H}=\sum_{i=1}^N \Big( -{\hbar^2\over 2 m} \nabla_i^2 
+ V_0({\bf r}_i) \Big) + 
{1\over 2} \sum_{ij=1}^N V({\bf r}_i,{\bf r}_j) \; ,
\eeq     
where $V_0({\bf r})$ is an external potential and $V({\bf r},{\bf r}')$  
is the interaction potential. 
In the mean-field approximation the totally symmetric 
many-particle wave-function of the Bose-Einstein condensate reads 
\beq 
\psi({\bf r}_1,...,{\bf r}_N,t) = \phi({\bf r}_1,t) \; ... 
\; \phi({\bf r}_N,t) \; , 
\eeq 
where $\phi ({\bf r},t)$ is the single-particle wave-function. 
Note that such factorization of the total wave-function is exact 
in the case of a non-interacting condensate. 
The quantum action of the system is then simply given by   
\beq
S_{GP}=N \int dt <\phi (t) | i\hbar 
{\partial \over \partial t} - {\hat h_s} |\phi (t) > \; ,
\eeq       
where 
\beq
{\hat h_s}= -{\hbar^2\over 2 m} \nabla^2 
+ V_0({\bf r}) + {1\over 2}(N-1) 
\int d^3{\bf r}' |\phi ({\bf r}')|^2 V({\bf r},{\bf r}') \; .
\eeq  
We call $S_{GP}$ the Gross-Pitaevskii (GP) action of the Bose condensate.  
By using the quantum least action principle  
we get the Euler-Lagrange equation 
\beq
i\hbar {\partial \over \partial t}\phi ({\bf r},t)= 
\Big[ -{\hbar^2\over 2m} \nabla^2 
+ V_0({\bf r}) + (N-1) 
\int d^3{\bf r}' V({\bf r},{\bf r}') |\phi ({\bf r}',t)|^2 
\Big] \phi ({\bf r},t)  \; , 
\eeq
which is an integro-differential nonlinear Schr\"odinger equation. 
Such equation and the effect of a finite-range interaction have 
been analyzed only by few authors (see, for example, Ref. 14-16). 
In fact, at low energies, it is possible to substitute 
the true interaction with a pseudo-potential$^{7}$  
\beq
V({\bf r},{\bf r}') = B \; \delta^3 ({\bf r}-{\bf r}') \; ,
\eeq
where $B={4\pi \hbar^2 a_s/m}$ is the scattering amplitude and $a_s$ 
the s-wave scattering length. In this way one obtains 
the so-called time-dependent GP equation$^{5,6}$ 
\beq
i\hbar {\partial \over \partial t}\phi ({\bf r},t)= 
\Big[ -{\hbar^2\over 2m} \nabla^2 
+ V_0({\bf r}) + B (N-1) |\phi ({\bf r},t)|^2 \Big] \phi ({\bf r},t)  \; ,
\eeq  
that is the starting point of many calculations.$^{4}$ 
Note that the GP equation is accurate to describe the condensate 
of weakly-interacting bosons only 
near zero temperature, where thermal excitations can be neglected. 
At finite temperature, one must consider a generalized GP equation plus 
the Bogoliubov equations, which describe the quasi-particle 
elementary excitations of the condensate.$^{7}$

\section{Trial Wave-Function of the Condensate}
\par
In this section we use a Gaussian trial wave-function to 
describe the ground-state of the Bose condensate and to minimize 
the GP action. 
\par
We consider a triaxially asymmetric harmonic trapping potential of
the form 
\begin{equation}
V_0\left({\bf r}\right)=
{1\over 2}m\omega_0^2\left(\omega_1^2 x^2+
\omega_2^2 y^2+\omega_3^2 z^2\right)\;,
\end{equation}
where $\omega_i$ ($i=1,2,3$) are adimensional constants proportional
to the spring constants of the potential along the three axes. 
In the rest of this section we write the lengths in units 
of the characteristic length of the 
trap $a_0=\sqrt{\hbar /(m\omega_0)}$, 
the time in units $\omega_0^{-1}$, the action in units $\hbar$ and 
the energy in units $\hbar \omega_0$. 
By using the delta pseudo-potential to model 
the interaction between particles, the GP action 
is explicitly given by 
\beq
S_{GP} = N \int dt \; d^3{\bf r}\;  
\phi^*({\bf r},t) 
\Big[ 
i {\partial \over \partial t} +{1\over 2} \nabla^2 
- V_0({\bf r}) - {1 \over 2} B (N-1) |\phi ({\bf r})|^2  
\Big] \phi ({\bf r},t) \; ,
\eeq             
where $B=4\pi a_s/a_0$ in our units. 
\par 
We want minimize this GP action $S_{GP}$ by choosing 
an appropriate single-particle trial wave-function.  
A good choice is the following 
\beq 
\phi\left({\bf r},t\right)= 
\Big[ {1\over \pi^3 {\sigma}_1^2(t) {\sigma}_2^2(t) 
{\sigma}_3^2(t)} \Big]^{1/4} 
\prod_{i=1,2,3} 
\exp\left\{-\frac{x_i^2}{2 {\sigma}_i^2(t)}
+i \beta_i(t) x_i^2 \right\}\;,
\eeq
with $(x_1,x_2,x_3)\equiv(x,y,z)$. $\sigma_i$ and $\beta_i$ are 
the time-dependent variational parameters. The $\sigma_i$ are 
the widths of the condensate in the three axial directions. 
Note that, in order to describe the time evolution of the variational 
function, the phase factor $i \beta_i(t) x_i^2$ is needed.$^{11,12}$ 
The choice of a Gaussian shape for the condensate 
is well justified in the limit of weak interatomic coupling, 
because the exact ground-state of the linear Schr\"odinger 
equation with harmonic potential is a Gaussian. Moreover, 
for the description of the collective
dynamics of Bose-Einstein condensates, 
it has already been shown that the variational technique based 
on Gaussian trial functions leads to
reliable results even in the large condensate number limit.$^{15-17}$  
\par 
By inserting the trial wave-function, 
after spatial integration, the GP action becomes 
\beq
S_{GP}= N \int dt \; L({\dot \beta_i},\beta_i, \sigma_i) \; ,
\eeq
where the effective Lagrangian is given by 
\beq
L=-{1\over 2} \Big[ \sum_{i=1}^3 
\big( {\dot \beta_i}\sigma_i^2 + {1\over 2 \sigma_i^2} + 
2 \sigma_i^2 \beta_i^2 + {1\over 2} \omega_i^2 \sigma_i^2 \big) 
+ {{\tilde g} \over \sigma_1\sigma_2\sigma_3} \Big] \; ,
\eeq
with $g=(2/\pi)^{1/2}(a_s/a_0)$ and ${\tilde g}=g(N-1)$. 
Again, by imposing the least action principle, we find 
six Euler-Lagrange equations
\beq
\beta_i = -{{\dot \sigma_i}\over 2 \sigma_i} 
\; , \;\;\;\;\;\;\;\;  
{\ddot \sigma_i}+\omega_i^2 \sigma_i = {1\over \sigma_i^3} + 
{{\tilde g} \over \sigma_1 \sigma_2 \sigma_3 \sigma_i} \; , 
\eeq
with $i=1,2,3$. Note that the time dependence of 
$\beta_i$ is fully determined by that of $\sigma_i$. Moreover, 
for ${\tilde g}=0$ the six equations are exact solutions 
of the time-dependent GP equation. 
The last three differential equations are the classical 
equations of motion of a system with coordinates $\sigma_i$ and
total energy per particle given by 
\beq 
{E\over N} = {1\over 2}
\Big[ \sum_{i=1}^3{1\over 2}\dot{\sigma}_i^2 \Big] + 
U(\sigma_1,\sigma_2,\sigma_3) \; , 
\eeq
where the potential energy is given by 
\beq
U(\sigma_1,\sigma_2,\sigma_3)= {1\over 2} \Big[ 
\sum_{i=1}^3 
\big({1\over 2}\omega_i^2\sigma_i^2 + {1\over 2\sigma_i^2}\big)   
+{{\tilde g}\over \sigma_1\sigma_2\sigma_3} \Big] \; . 
\eeq 
We use the previous equations to calculate the ground-state energy 
and the low-energy excitations of the condensate. 
\par
The ground-state energy per particle of the Bose 
condensate is simply $E^{(0)}/N=U({\vec \sigma}^*)$, where 
${\vec \sigma^* }=(\sigma_1^*,\sigma_2^*,\sigma_3^*)$ 
is the the minimum of the effective potential energy $U$. 
This equilibrium point is identified by the following equations:
\beq
\omega_1^2\sigma_1^4 -{\tilde g}\frac{\sigma_1}{\sigma_2\sigma_3} =1\; ,
\;\;\;
\omega_2^2\sigma_2^4 -{\tilde g}\frac{\sigma_2}{\sigma_1\sigma_3} =1\; ,
\;\;\;
\omega_3^2\sigma_3^4 -{\tilde g}\frac{\sigma_3}{\sigma_1\sigma_2} =1\; . 
\label{minimum}
\end{equation}
\par 
The low-energy collective excitations of the condensate are 
the small oscillations of variables $\sigma_i$'s around the equilibrium 
point. The calculation of the normal mode frequencies for the motion of the 
condensate is reduced to an eigenvalue problem for the 
Hessian matrix $\Lambda$, 
given by  
\begin{equation}
\Lambda_{ij}=
\left.\frac{\partial^2U}{\partial\sigma_i\partial\sigma_j}
\right|_{{\vec \sigma}={\vec \sigma}^*}\; .
\end{equation} 
\par 
In the case ${\tilde g}=0$, the stationary solution is simply 
$\sigma_i^*=\omega_i^{1/2}$ $(i=1,2,3)$ and 
the energy per particle of the non-interacting condensate is given by 
${E^{(0)}/N}=(\omega_1+\omega_2 + \omega_3)/2$. 
The Hessian matrix $\Lambda$ of the potential energy $U$ is diagonal 
and the frequency modes are simply $\Omega_i=2 \omega_i$ $(i=1,2,3)$. 
\par 
The general solution of Eq.~(\ref{minimum}) for the stationary state 
and of the eigenvalue problem for the collective mode frequencies can 
be obtained numerically,$^{17}$ but analytical results are available 
in the Thomas-Fermi limit. 
In such limit, i.e. when ${\tilde g}\gg 1$, 
the right-hand side in Eq.~(\ref{minimum}), 
related to the kinetic pressure, can be 
neglected and the following expression for the
equilibrium point is obtained 
\beq
\sigma_1^*=\Big({{\tilde g} \; \omega_2 \; \omega_3
\over \omega_1^4} \Big)^{1/5} \; , 
\;\;\;\;\; 
\sigma_2^*=\Big({{\tilde g} \; \omega_1 \; \omega_3 
\over \omega_2^4} \Big)^{1/5} \; ,
\;\;\;\;\; 
\sigma_3^*=\Big({{\tilde g} \; \omega_1 \; \omega_2 
\over \omega_3^4} \Big)^{1/5} \; .
\eeq
The energy per particle of the ground-state results
\beq 
{E^{(0)}\over N} = {5\over 4} 
\Big( {\tilde g} \; \omega_1 \; \omega_2 \; \omega_3 \Big)^{2/5} \; , 
\eeq 
and the Hessian matrix $\Lambda$ is given by
\beq
\Lambda = 
\left( \begin{array}{ccc}  
       3 \; \omega_1^2   & \omega_1 \omega_2 & \omega_1 \omega_3 \\ 
       \omega_1 \omega_2 & 3 \; \omega_2^2   & \omega_2 \omega_3 \\ 
       \omega_1 \omega_3 & \omega_2 \omega_3 & 3 \; \omega_3^2 \\ 
\end{array}  \right) \; .
\eeq 
Note that ${\tilde g}$ does not appear in the matrix. 
The eigenfrequencies $\Omega$ 
of the collective motion are found as the solutions of the equation$^{14}$  
\beq 
\Omega^6
-3\left(\omega_1^2+\omega_2^2+\omega_3^2\right)\Omega^4+
8\left(\omega_1^2\; \omega_2^2+\omega_1^2\; \omega_3^2
+\omega_2^2\; \omega_3^2 \right)\Omega^2
-20\; \omega_1^2\; \omega_2^2\; \omega_3^2=0 \; .
\eeq 
This equation has been also obtained by Stringari$^{4}$ using a hydrodynamical 
approach, which is effective in the Thomas-Fermi limit.  
\par
For an axially symmetric trap, where 
$\omega_1=\omega_2=\omega_{\bot} \neq \omega_3$, 
the previous equations give 
$\sigma_1^* =\sigma_2^* =({\tilde g}\omega_3/\omega_{\bot}^3)^{1/5}$, 
$\sigma_3^*=({\tilde g}\omega_{\bot}/\omega_3^3)^{1/5}$, 
$E^{(0)}/N=(5/4){\tilde g}^{2/5} \omega_{\bot}^{4/5}\omega_3^{2/5}$, and 
\beq 
\Omega_{1,2}=\sqrt{2\omega_{\bot}^2 + {3\over 2}\omega_3^2 \pm 
{1\over 2}\sqrt{16 \omega_{\bot}^4 + 9\omega_3^4 -
16\omega_{\bot}^2\omega_3^2}} \; ,
\;\;\;\;
\Omega_3 =\sqrt{2} \; \omega_{\bot} \; .
\eeq 
Because of the axial symmetry of the trap, the angular momentum 
along the $z$ axis is conserved and one can thus label the modes 
by angular ($l$) and azimuthal ($m_z$) quantum numbers. 
One finds$^{4,11}$ $m_z=0$ for the coupled monopole ($l=0$) 
and quadrupole ($l=2$) modes 
$\Omega_{1,2}$, and $|m_z|=2$ for the quadrupole ($l=2$) 
mode $\Omega_3$. Note that the experimental results obtained 
on sodium vapors at MIT are in excellent agreement with these 
theoretical values.$^{18}$  
\par
For an isotropic harmonic trap with frequency $\omega$, we obtain 
$\sigma_i^* =({\tilde g}/\omega^2)^{1/5}$ ($i=1,2,3$), 
$E^{(0)}/N=(5/4){\tilde g}^{2/5} \omega^{6/5}$, and 
$\Omega_{1,2}=\sqrt{5}\; \omega$, $\Omega_3 = \sqrt{2} \; \omega$. 
Here, differently from the axially symmetric case, the frequencies 
do not depend on $m_z$. In particular, 
$\sqrt{5}\omega$ is the monopole oscillation, also called 
breathing mode, characterized by radial quantum number 
$n_r=1$ and angular quantum number $l=0$. 
Instead $\sqrt{2}\omega$ is the quadrupole ($l=2$) surface ($n_r=0$) 
oscillation. 
\par 
It is worth noting that the variational formalism 
makes it possible to investigate the case of negative scattering length. 
In this case, the large $N$ limit of the Thomas-Fermi approximation 
does not apply because 
the condensate number cannot exceed a critical value. 
Let us consider again the isotropic trap with frequency 
$\omega$. For the ground-state, the number of particles 
satisfies the following equation 
\beq 
N = 1+ {1\over g} (\omega^2 \sigma_0^5-\sigma_0) \; , 
\eeq 
where $\sigma_0 =\sigma_i^*$ ($i=1,2,3$) is mean radius of the 
condensate, that is the same in the three axial directions. 
It follows that, for $g>0$, with $N=1$ we have $\sigma_0 =1$ 
and with $N\to \infty$ then $\sigma_0\to \infty$. 
Instead, for $g<0$, the mean radius $\sigma_0$ decreases 
by increasing the number of bosons 
$N$ to a critical radius $\sigma_0^c =(5\omega^2)^{-1/4}$ 
with a critical number of bosons given by 
$N^c=1+|g|^{-1}\omega^{-1/2}(5^{-1/4}-5^{-5/4})$. 
Moreover, one finds a simple expression for the monopole ($l=0$) 
frequency of surface oscillation of the condensate, namely$^{16}$ 
\beq 
\Omega = \Big[5\omega^2 -\sigma_0^{-4}\Big]^{1/2} \; . 
\eeq 
We have seen that $\sigma_0=1$ for $N=1$ or $g=0$. Moreover, 
when $N>1$, it is $\sigma_0>0$ for $g>0$ 
and $\sigma_0^c<\sigma < 1$ for $g<0$. 
Regarding the frequency $\Omega$, 
one easily verifies that $\Omega \to \sqrt{5}\omega$ for 
$\sigma_0 \to \infty$ and $\Omega \to 0$ for $\sigma_0\to \sigma_0^c$. 
Note that, for the triaxial trap, at the critical point 
the two highest frequencies diverge and the 
lowest one falls to zero.$^{17}$ 
\par 
It is important to observe that it is possible to include 
in the Gaussian trial wave-function other three  parameters 
describing the position of the center of mass of the condensate. 
Due to the harmonic confinement, the motion of the center 
of mass (dipole mode) is periodic and the frequencies of 
oscillation are equal to those of the harmonic trap.$^{11,12}$ 

\section{Trial Wave-Function of Vortex States}
\par
Let us consider states having a vortex line 
along the $z$ axis and all bosons flowing around it with 
quantized circulation. The observation of these states would be 
a signature of macroscopic phase coherence of trapped BEC. 
The time-dependent variational approach can be used 
to describe also these vortex states of the condensate in a 
axially symmetric trap 
\beq
V_0({\bf r})={1\over 2} \left( \omega_{\bot}^2 r^2 + 
\omega_3^2 z^2 \right) \; ,
\eeq
where $r=\sqrt{x^2+y^2}$, $\theta=$arctg$(y/x)$ and $z$ are the 
cylindrical coordinates. 
\par
The trial wave function of the vortex states of the condensate 
can be chosen as 
\beq 
\phi_k \left({\bf r},t \right)= 
\Big[{1\over k! \pi^3 {\sigma_{\bot}}^{4k+4}(t)
{\sigma}_3^2(t) }\Big]^{1/4}
r^k e^{i k \theta} 
\exp\left\{-\frac{r^2}{2 \sigma_{\bot}^2(t)}-{z^2\over 2 \sigma_3^2(t)} 
+i \beta_{\bot}(t) r^2 + i \beta_3(t) z^2 \right\}\;,
\eeq
where $k$ is the vortex quantum number and 
$\sigma_{\bot}$, $\sigma_3$, $\beta_{\bot}$ and $\beta_3$ are 
the time-dependent variational parameters.
This trial wave-function describes exactly the 
vortex states of a non-interacting gas of bosons in the harmonic trap. 
\par 
By following the same procedure of the previous section, 
namely by inserting the trial wave-function in the GP action, 
and after spatial integration, one gets 
\beq
S_{GP}= N \int dt \; L({\dot \beta_{\bot}},{\dot \beta_3},
\beta_{\bot} ,\beta_3, \sigma_{\bot}, \sigma_3) \; ,
\eeq
where the effective Lagrangian is given by 
$$
L=-{1\over 2} \Big[ (k+1)\big( {\dot \beta_{\bot}}\sigma_{\bot}^2 + 
{1\over 2 \sigma_{\bot}^2} + 2 \sigma_{\bot}^2 \beta^2 + {1\over 2} 
\omega_{\bot}^2 \sigma^2  \big)+
$$
\beq
+\big( {\dot \beta_3}\sigma_3^2 + {1\over 2 \sigma_3^2} + 
2 \sigma_3^2 \beta_3^2 + {1\over 2} 
\omega_3^2 \sigma_3^2  \big)    
+ {{\tilde g}(k) \over \sigma_{\bot}^2\sigma_3} \Big] \; ,
\eeq
with $g=(2/\pi)^{1/2}(a_s/a_0)$ and 
${\tilde g}(k)=g(N-1)(2k)!/(2^{2k}k!^2)$. 
The four Euler-Lagrange equations of the system are given by 
\beq 
\beta_{\bot} = -{{\dot \sigma_{\bot}}\over 2 \sigma_{\bot}} 
\; , \;\;\;\;\;\;\;\;\;\; 
\beta_3 = -{{\dot \sigma_3}\over 2 \sigma_3} \; , 
\eeq
\beq 
(k+1){\ddot \sigma_{\bot}}+(k+1)\omega_{\bot}^2 \sigma_{\bot}
= {(k+1)\over \sigma_{\bot}^3} + 
{{\tilde g}(k) \over \sigma_{\bot}^3 \sigma_3} \; , 
\;\;\; \;\;\;
{\ddot \sigma_3}+\omega_3^2 \sigma_3 = {1\over \sigma_3^3} + 
{{\tilde g}(k) \over \sigma_{\bot}^2 \sigma_3^2} \; . 
\eeq
The time dependence of $\beta_{\bot}$ and $\beta_3$ 
is fully determined by that of $\sigma_{\bot}$ and $\sigma_3$; 
moreover, for ${\tilde g}=0$ (non-interacting condensed vortex) 
these four equations are exact. 
The last two differential equations correspond to the classical 
equations of motion of a system with energy per particle 
\beq 
{E_k\over N} = {1\over 2} \Big[ 
(k+1)\dot{\sigma}_{\bot}^2 + {1\over 2}\dot{\sigma}_3^3 \Big] 
+ U(\sigma_{\bot},\sigma_3) \; , 
\eeq
where the potential energy is given by 
\beq
U(\sigma_{\bot},\sigma_3)= {1\over 2} \Big[
(k+1)\omega_{\bot}^2\sigma_{\bot}^2 + 
{1\over 2}\omega_3^2\sigma_3^2 + 
{(k+1)\over \sigma_{\bot}^2}+ {1\over 2\sigma_3^2} 
+{{\tilde g}(k)\over \sigma_{\bot}^2 \sigma_3} \Big] \; . 
\eeq 
The equilibrium point $(\sigma_{\bot}^*,\sigma_3^*)$, 
corresponding to the minimum 
of the effective potential energy, is given by the following equations:
\beq
(k+1)\omega_{\bot}^2\sigma_{\bot}^4 
-{\tilde g}(k){1\over \sigma_3} =(k+1)\; ,
\;\;\; \;\;\;\;
\omega_3^2\sigma_3^4 -{\tilde g}(k)\frac{\sigma_3}{\sigma_{\bot}^2} =1\; .
\eeq
\par 
In the non-interacting case $({\tilde g}=0)$ 
the stationary solution is simply 
$\sigma_{\bot}^*=\omega_{\bot}^{1/2}$ and  
$\sigma_3^*=\omega_3^{1/2}$. 
The stationary energy of the non-interacting vortex state 
is exactly given by 
\beq
{E_k^{(0)}\over N} = (k+1)\omega_{\bot}+{1\over 2}\omega_3 \; .
\eeq   
It is important to observe that the energy grows only 
linearly with the vortex quantum number $k$, in contrast 
to the homogenous case$^{7}$ where the vortex energy grows as $k^2$. 
In addition, the two ($m_z=0$) frequency modes are  
$\Omega_1=2 \omega_{\bot}$ and $\Omega_2=2 \omega_3$ 
and do not depend on the vortex quantum number $k$. 
\par 
In the Thomas-Fermi limit (${\tilde g}>>1$), the analytic 
expression of the equilibrium point reads  
\beq
\sigma_{\bot}^*=\left[ {\tilde g}(k) \; \omega_3 
\over (k+1)^{3/2}\omega_{\bot}^3\right]^{1/5} \; , \;\;\; \;\;\;  
\sigma_3^*=\left[{{\tilde g}(k) \; (k+1)\; \omega_{\bot}^2 
\over \omega_3^4} \right]^{1/5} \; . 
\eeq 
The stationary energy of the vortex state results 
\beq
{E_k^{(0)}\over N}= {5\over 4} \Big( (k+1) \; {\tilde g}(k) 
\; \omega_{\bot}^2 \; 
\omega_3 \Big)^{2/5} \; .
\eeq 
From this formula it is easy to calculate the critical frequency 
${\bar \Omega}_c$ at which a vortex can be produced. 
One has to compare the energy of a vortex state in frame rotating 
with angular frequency ${\bar \Omega}$, that is 
$E^{(0)}_k-{\bar \Omega} L_z$, with 
the energy $E^0$ of the ground state (with no vortices).$^{7}$ 
Since the angular momentum per particle is $\hbar k$, the critical 
frequency is given by $\hbar {\bar \Omega}_c =(E_k^{(0)}/N - E^{(0)}/N)/k$.  
By using our previous formula, we get  
\beq
{\bar \Omega}_c = {5\over 4} k^{-1} \omega_{\bot}^{4/5} \omega_3^{2/5} 
\Big[ {\tilde g}(k)^{2/5} (k+1)^{2/5} - {\tilde g}(0)^{2/5} \Big] \; .  
\eeq 
In the presence of a vortex, 
for frequencies smaller than the critical frequency ${\bar \Omega}_c$, 
the system is not thermodynamically 
stable (global stability), but can be locally stable (metastability) 
against small deformations of the system. Rohksar$^{19}$ 
has shown that the vortex state is locally unstable 
for ${\bar \Omega}=0$. Actually, in a harmonic trap, 
there is local instability for $0\leq {\bar \Omega} < {\bar \Omega}_s$, where 
the frequency ${\bar \Omega}_s$ (${\bar \Omega}_s < {\bar \Omega}_c$) 
can be estimated numerically.$^{20}$ 
On the other hand, we have recently shown that, 
in toroidal traps, vortices can be locally stable also 
in absence of external forced rotation.$^{21}$ 
\par
It is important to observe that, in our present treatment, 
the condensate has a unique vortex line, which is fixed along 
the $z$ axis. Butts and Rokhsar$^{22}$ have numerically shown that 
increasing the rotational frequency ${\bar \Omega}$ well beyond 
${\bar \Omega}_c$ will favour the creation of more complex vortical 
configurations, associated with the occurrence of $2$ or more vortex lines. 
\par
To calculate the collective oscillations 
in the Thomas-Fermi limit, we follow the same 
procedure of the previous section. The Hessian matrix $\Lambda$ of 
the potential energy $U$ at the equilibrium point is given by
\beq 
\Lambda = 
\left( \begin{array}{cc}  
8\; (k+1)\; \omega_{\bot}^2  & 2\; (k+1)^{1/2}\; \omega_{\bot}\; \omega_3  \\  
2\; (k+1)^{1/2}\; \omega_{\bot}\; \omega_3 & 3\; \omega_3^2  \\ 
\end{array}  \right) \; .
\eeq 
Again the ${\tilde g}(k)$ dependence drops out. 
Moreover, the mass matrix of the kinetic energy is not the identity 
matrix but instead  
\beq
M= 
\left( \begin{array}{cc}  
     2 \; (k+1) & 0  \\  
     0       & 1  \\ 
\end{array}  \right) \; .
\eeq
The eigenfrequencies $\Omega$ 
of the collective motion are found as the solutions of the equation 
$|\Lambda - \Omega^2 M| = 0$, which gives the $m_z=0$ frequencies
\beq
\Omega_{1,2}=\sqrt{2\omega_{\bot}^2 + {3\over 2}\omega_3^2 \pm 
{1\over 2}\sqrt{16 \omega_{\bot}^4 + 9\omega_3^4 -
16\omega_{\bot}^2\omega_3^2}} \; . 
\eeq
This is an interesting result: The $m_z=0$ frequencies 
of a harmonically trapped condensate do not depend on 
the vortex quantum number $k$. Note that Zambelli and Stringari$^{23}$ 
have shown, by using a sum-rule approach, that the 
$m_z\neq 0$ collective frequencies do instead depend on $k$. 
Nevertheless, all the modes are $k$ sensitive because of 
the amplitudes, which are given by $<r^2>^{1/2}=(k+1)^{1/2}\sigma_{\bot}$ 
and $<z^2>^{1/2}=\sigma_3/\sqrt{2}$.  
\par 
Finally, we study an isotropic trap with 
frequency $\omega$. We have $\sigma^*_{\bot}=\sigma^*_3$ 
only for ${\tilde g}=0$. In order to obtain some analytic 
results also in the case of negative scattering length, 
we consider the vortex trial wave-function with 
a unique variational parameter $\sigma$, i.e. 
we put $\sigma_{\bot}=\sigma_3=\sigma$. 
For the ground-state, the number of particles 
satisfies the equation $N = 1+g(k)^{-1} 
(\omega^2\sigma_0^5-\sigma_0) (1+{2/3}k)$, 
where $\sigma_0$ is mean radius of the condensate in the 
stationary configuration and $g(k)=g(2k)!/(2^{2k}k!^2)$. 
It follows that for $g<0$, the critical number of bosons 
for the collapse of the wave-function is given 
by $N^c=1+|g(k)|^{-1}\omega^{-1/2}(5^{-1/4}-5^{-5/4})(1+2/3k)$. 
Thus, compared with the condensed state without vortices, 
the critical number of trapped atoms in the vortex state increases. 
This variational result is in full agreement 
with the $k=1$ numerical results of Dalfovo and Stringari.$^{24}$ 
            
\section{Conclusions}

We have studied a Bose-Einstein condensate trapped 
in a harmonic potential by using a variational method based on 
the minimization of the Gross-Pitaevskii action with a Gaussian ansatz 
for the shape of the macroscopic wave function of the
condensate. Our analysis has been applied to the whole range of 
coupling strengths between particles (attractive and 
repulsive interaction) to study both the ground-state 
and vortex states. In particular, for vortex states, 
we have determined the critical rotational 
frequency at which a single vortex becomes globally stable. 
Moreover, we have calculated the collective oscillations 
and the critical number of bosons for the collapse of the condensate 
in the case of attractive interaction. 
Nowadays it is experimentally possible to trap and condense a very large 
number of alkali atoms (more than $10^{9}$) and to obtain 
a less diluite system. Thus, the study 
of the time-dependent variational approach 
without the independent-particle approximation 
begins to appear an interesting problem also for Bose gases 
in an external potential (the case of a superfluid film containing a vortex 
has been discussed in Ref. 9 and 10). 
This will be one of our future projects. 

\section*{Acknowledgments} 

This work has been supported by INFM through a Research Advanced Project 
on Bose-Einstein Condensation. The author thanks L. Reatto and 
A. Parola and D. Pini for useful discussions. 

\newpage

\section*{References}

\begin{description}

\item{\ 1.} M.H. Anderson {\it et al.}, Science {\bf 269}, 198 (1995). 
 
\item{\ 2.} C.C. Bradley {\it et al.}, 
Phys. Rev. Lett. {\bf 75}, 1687 (1995). 

\item{\ 3.} K.B. Davis {\it et al.}, 
Phys. Rev. Lett. {\bf 75}, 3969 (1995). 

\item{\ 4.} F. Dalfovo, S. Giorgini, L.P. Pitaevskii, and S. Stringari, 
Rev. Mod. Phys. {\bf 71}, N.3 (1999). 

\item{\ 5.} E.P. Gross, Nuovo Cimento {\bf 20}, 454 (1961). 

\item{\ 6.} L.P. Pitaevskii, Zh. Eksp. Teor. Fiz. {\bf 40}, 
646 (1961) [English Transl. Sov. Phys. JETP {\bf 13}, 451 (1961)].

\item{\ 7.} A. Fetter and J. Walecka, {\it Quantum Theory 
of Many Particle Systems} (McGraw-Hill, New York, 1971). 

\item{\ 8.} G.D. Mahan, {\it Many-Particle Physics} 
(Plenum Press, New York, 1990). 

\item{\ 9.} Q. Niu, P. Ao, and D.J. Thouless, 
Phys. Rev. Lett. {\bf 72}, 1706 (1994).  

\item{\ 10.} D. Demircan, P. Ao, and Q. Niu, 
Phys. Rev. B {\bf 54}, 10027 (1996).  

\item{\ 11.} V.M. P\'erez-Garcia, H. Michinel, J.I. Cirac, M. Lewenstein,
and P. Zoller, Phys. Rev. Lett. {\bf 77}, 5320 (1996) 

\item{\ 12.} V.M. P\'erez-Garcia, H. Michinel, J.I. Cirac, M. Lewenstein,
and P. Zoller, Phys. Rev. A {\bf 56} 1424 (1997). 

\item{\ 13.} L. Salasnich, Mod. Phys. Lett. B {\bf 11}, 1249 (1997). 

\item{\ 14.} L. Salasnich, Mod. Phys. Lett. B {\bf 12}, 649 (1998). 

\item{\ 15.} A. Parola, L. Salasnich, and L. Reatto, 
Phys. Rev. A {\bf 57}, R3180 (1998). 

\item{\ 16.} L. Reatto, A. Parola, and L. Salasnich, 
J. Low Temp. Phys. {\bf 113}, 195 (1998). 

\item{\ 17.} E. Cerboneschi, R. Mannella, E. Arimondo, and L. Salasnich, 
Phys. Lett. A {\bf 249}, 495 (1998). 

\item{\ 18.} D.M. Stamper-Kurn {\it et al.}, cond-mat/9801262. 

\item{\ 19.} D.S. Rokhsar, Phys. Rev. Lett. {\bf 79}, 2164 (1997). 

\item{\ 20.} A.A. Svidzinsky and A.L. Fetter, cond-mat/9811348. 
 
\item{\ 21.} L. Salasnich, A. Parola, and L. Reatto, 
Phys. Rev. A {\bf 59}, 2990 (1999). 
 
\item{\ 22.} D. A. Butts and R. Rokhsar, Nature {\bf 397}, 327 (1999). 

\item{\ 23.} Zambelli and Stringari, 
Phys. Rev. Lett. {\bf 81}, 1754 (1998). 

\item{\ 24.} F. Dalfovo and S. Stringari, 
Phys. Rev. A {\bf 53}, 2477 (1996). 
             
\end{description}
                            
\end{document}